\documentclass[conference]{IEEEtran}
\IEEEoverridecommandlockouts

\usepackage{fancyhdr, epsfig, epsf, amsthm, amsmath, amssymb, amsfonts, subfigure, color}
\usepackage{graphicx}
\usepackage{textcomp}
\usepackage{xcolor}
\usepackage{ragged2e}
\usepackage{algorithm}
\usepackage{dsfont}
\usepackage{enumerate}
\usepackage{threeparttable}
\usepackage{soul}
\usepackage{algpseudocode}
\usepackage[thinc]{esdiff}
\usepackage{cancel}
\usepackage{multirow}
\usepackage{adjustbox}
\usepackage[noadjust]{cite}

\newtheorem{proposition}{Proposition}

\def\BibTeX{{\rm B\kern-.05em{\sc i\kern-.025em b}\kern-.08em
    T\kern-.1667em\lower.7ex\hbox{E}\kern-.125emX}}
\begin{document}

\title{\LARGE Enabling AI Quality Control via Feature Hierarchical Edge Inference 
}
\author{\IEEEauthorblockN{Jinhyuk Choi$^*$, Seong-Lyun Kim$^*$, Seung-Woo Ko$^\S$}\\

\IEEEauthorblockA{$^*$School of EEE, {Yonsei University}, Seoul, Korea, email: \{jh.choi, slkim\}@ramo.yonsei.ac.kr}
\IEEEauthorblockA{$^\S$Dept. of Smart Mobility Eng., {Inha University}, Incheon, Korea, email: swko@inha.ac.kr}
}

\maketitle

\begin{abstract}
With the rise of edge computing, various AI services are expected to be available at a mobile side  through the inference based on \emph{deep neural network} (DNN) operated at the network edge, called \emph{edge inference} (EI). On the other hand, the resulting AI quality (e.g., mean average precision in objective detection) has been regarded as a given factor, and AI quality control has yet to be explored despite its importance in addressing the diverse demands of different users. This work aims at tackling the issue by proposing a \emph{feature hierarchical EI} (FHEI), comprising {feature network}  and {inference network} deployed at an edge server and corresponding mobile,  respectively. Specifically, feature network is designed based on feature hierarchy, a one-directional feature dependency with a different scale. A higher scale feature requires more computation and communication loads while it provides a better AI quality. The tradeoff enables FHEI to control AI quality gradually w.r.t. communication and computation loads, leading to deriving a near-to-optimal solution to maximize multi-user AI quality under the constraints of uplink \& downlink transmissions and edge server and mobile computation capabilities. 
It is verified by extensive simulations that the proposed joint communication-and-computation control on FHEI architecture always outperforms several benchmarks by differentiating each user's AI quality depending on the communication and computation conditions.
\end{abstract}

\section{Introduction}
Due to the rapid advancement of \emph{artificial intelligence} (AI), a wide range of mobile services has been built on an AI framework, e.g., Tensorflow and Torch, to offer accurate and reliable results by inferring the most likely outcomes using a well-trained \emph{deep neural network} (DNN). Edge computing-based inference, shortly \emph{edge inference} (EI), is expected to be its crucial enabler such that an edge server nearby runs the DNN to execute mobiles' inference tasks \cite{xu2018nature}. 

In an early stage of EI research, an AI model is assumed to be indivisible and installed at either a mobile or an edge server with communication efficient offloading techniques (see e.g., \cite{Prior1} and \cite{Prior2}). With the recent rise of split learning \cite{distributed_learning_survey}, it is possible to divide the entire AI model into multiple sub-models, called \emph{model partitioning} (MP).
With MP,  mobiles and an edge server can cooperate to perform EI, thereby reducing the computation load and computation latency. For example, in \cite{Task_Orient_EI_MP}, mobiles are in charge of extracting and encoding task-specific features from raw data based on the well-known information bottleneck principle, while the edge server proceeds the remaining EI by receiving the encoded features. Prompted by BranchyNet proposed in \cite{Branchynet}, the concept of early exit is introduced in \cite{Prior7} that multiple branches in the DNN return different features used as inputs to the corresponding EI processes. The time required to reach each branch is different, allowing the system to determine one of them as an exit branch depending on the computation latency requirement.

\begin{figure}
\centering
\includegraphics[width=8.5cm]{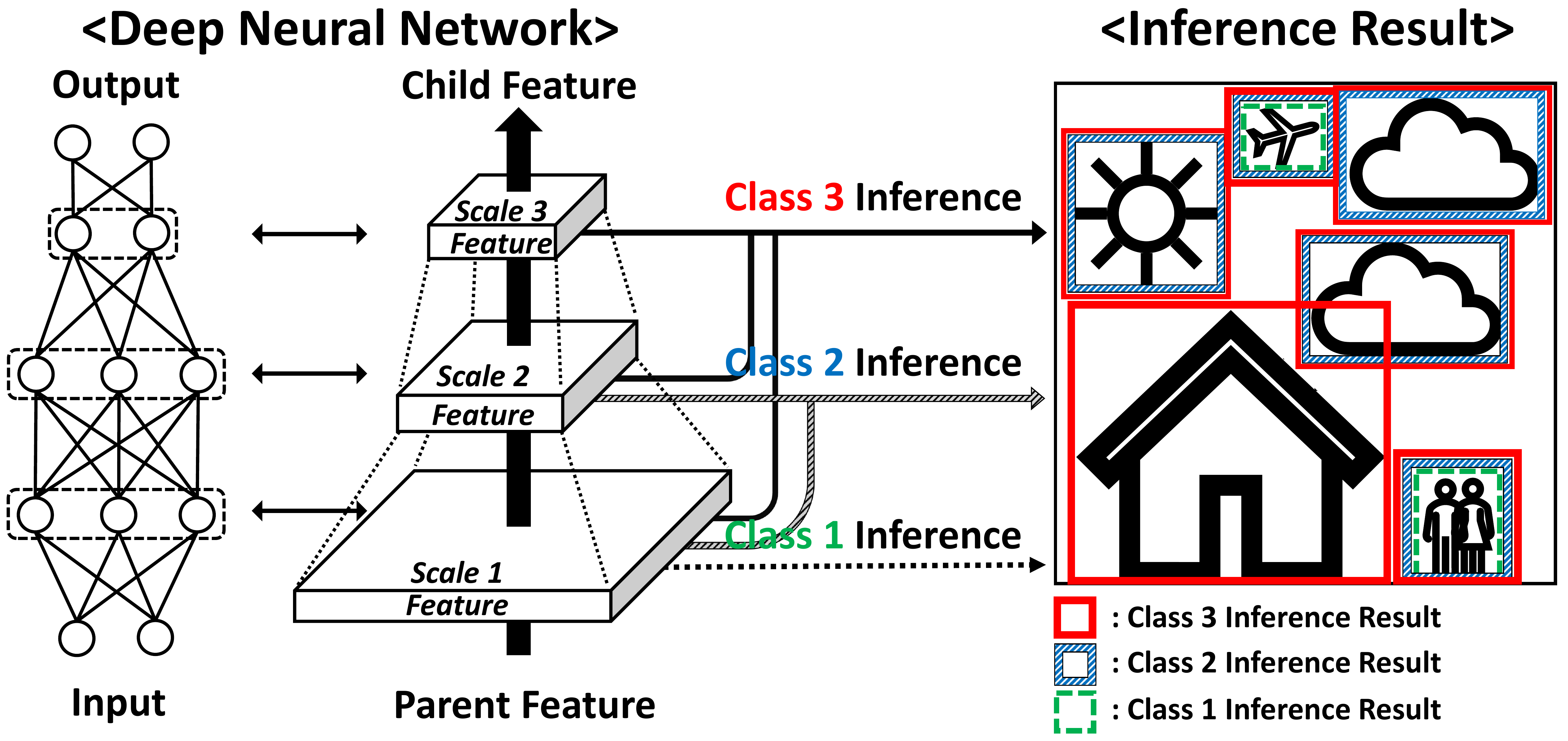}
\caption{Graphical representation of feature hierarchy of a DNN-based AI model.
A class-$k$ inference is defined as the inference using the features of scales $1$ to $k$ as inputs, achieving a more qualified result 
than a lower class~one.}\label{Fig:Feature_Hierachy}
\end{figure}

As aforementioned, most prior works on EI focuses on addressing the load balancing and latency issues, whereas the quality of AI has not been considered yet due to the following two reasons. First, the concerned DNN architecture returns an output with a single quality (e.g., \cite{Task_Orient_EI_MP}). Second, the result through a heavier computation does not guarantee a better quality (e.g., \cite{Prior7}), hindering an elastic control of AI quality on the concerned DNN architecture.

This work attempts to tackle the issue of controlling AI quality by exploiting feature hierarchy explained as follows. As shown in Fig. \ref{Fig:Feature_Hierachy}, each layer's activation pattern during a DNN operation corresponds to a feature vector extracted from the input data. It is well-known that the feature from a deeper layer can represent a larger scale of input data. Then, we define a feature of scale $k$ as the features extracted from the $k$-th layer. Due to a forward propagation process, the features of scales $(k-1)$ and $k$ have a hierarchical relation, 
transforming the former into the latter but not vice versa. The \emph{feature pyramid network} (FPN) proposed in \cite{FeaturePyramid} is a representative DNN architecture designed based on feature hierarchy, showing that when features of different scales are simultaneously used as inputs for an inference task, the resultant output has a better quality, e.g., many targets with different scales are well captured for object detection. On the other hand, extracting different scales of features requires a heavier computation load.

Inspired by the above trade-off, we propose a \emph{feature-hierarchical EI} (FHEI), which enables us to gradually control each mobile's AI quality. Specifically, we divide a feature hierarchy-based DNN into \emph{feature network} (FN) and \emph{inference network} (IN). Due to the heavy computation loads to extract multi-scale features, FN is installed at the edge server, while IN is located at each mobile to facilitate a user-customized service.
The edge server adjusts the degree of feature scale depending on the user's service quality demands under the constraints of its computation capability. Besides, FHEI requires not only uplink transmission to offload mobiles' local data to the edge server but also downlink transmission to return the extracted features to the corresponding mobiles. As a result, a joint radio-and-computation resource optimization is required to maximize sum AI quality, verified to achieve superior performance to several benchmarks.
\section{Feasibility Study on AI Quality Control}\label{Sec: Feasibility}
This section studies the feasibility of AI  quality control via the experiments explained below, leading to establishing the relation among multiple metrics with interesting insights.  

\begin{figure}[t!]
    \centering
        \subfigure[Computation load vs. Communication load]{
        \includegraphics[width=4cm]{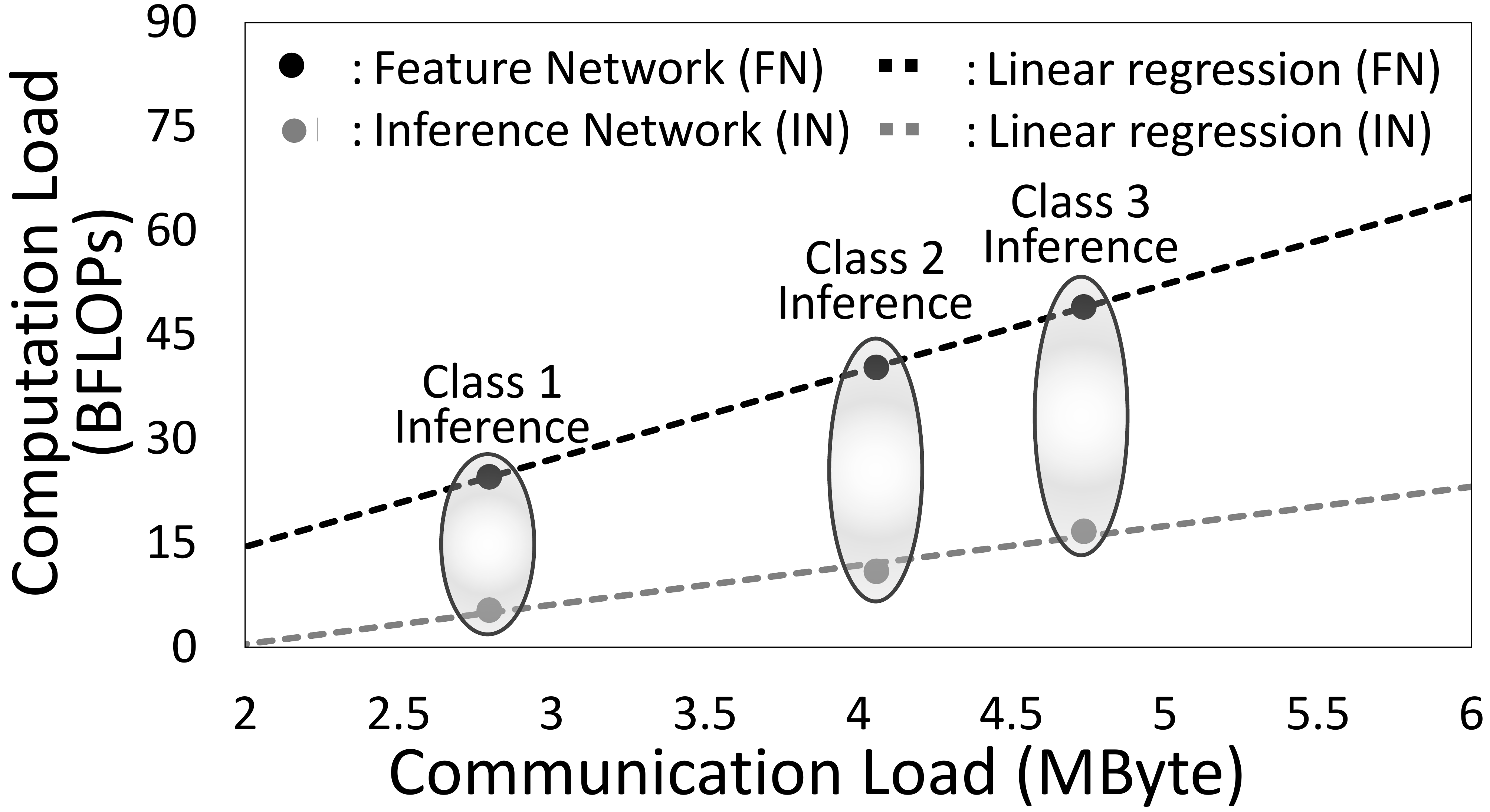}
        \label{Exp_Graph 1-1}
        }
        \subfigure[AI quality vs. Communication load]{
        \includegraphics[width=4cm]{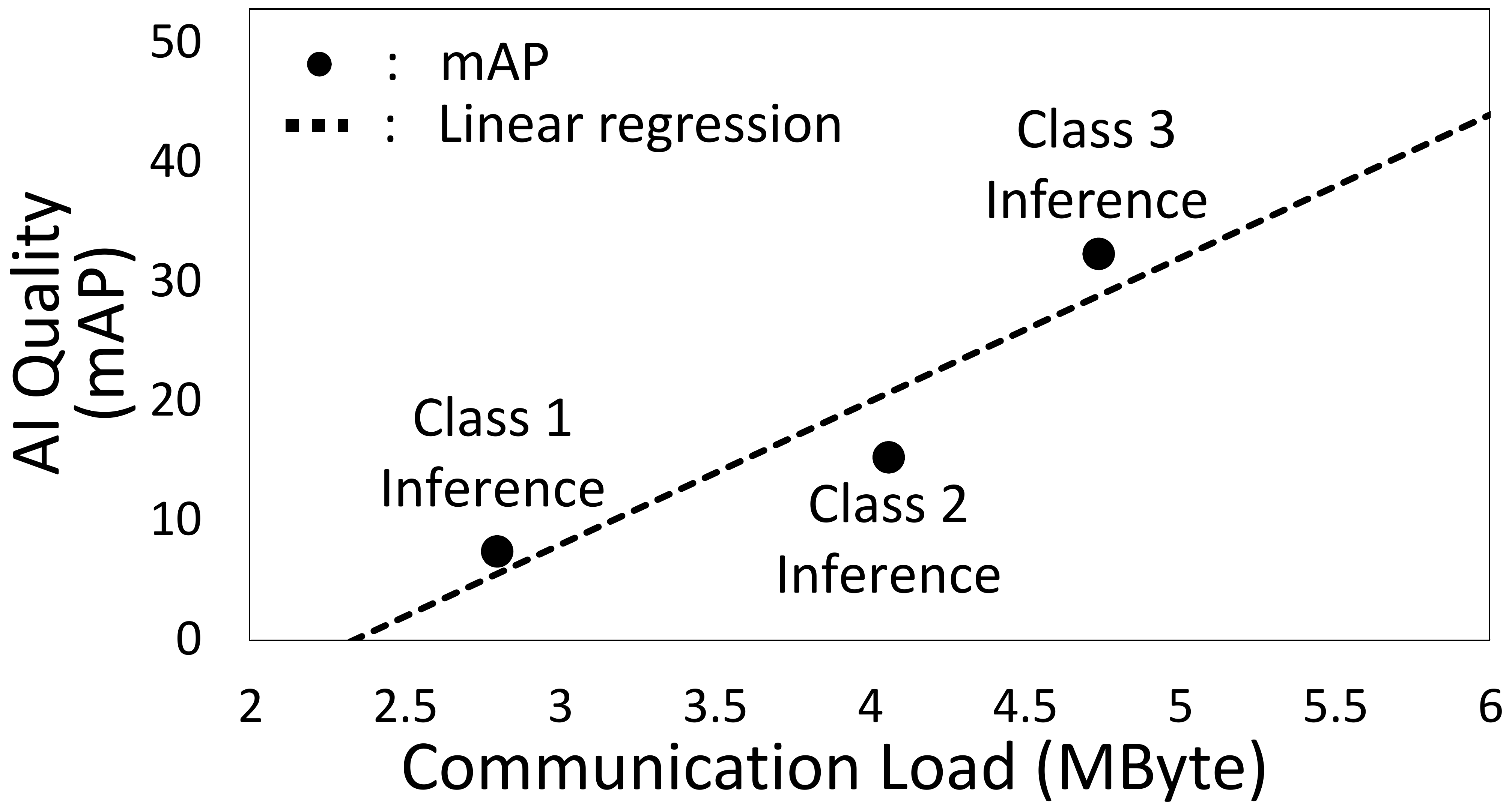}
        \label{Exp_Graph 1-2}
        }
     \caption{The relation among computation load, AI quality, and communication loads of feature hierarchy-based DNN with $3$ classes of inferences.}\label{Fig:Service_Quality_Validation}
\end{figure}

\subsection{Experiment Setting} 
We use the YOLO v3 for an object detection task, extracting multi-scale features based on FPN \cite{FeaturePyramid}\footnote{Various DNN structures built on feature hierarchy exist in the literature, such as U-Net \cite{U-Net} and a Laplacian pyramid \cite{LaplacianPyramid}. The experiments in the section are applicable to them, remaining as future work due to the page~limit.}. 
The concerned YOLO model is published in \cite{YOLO}, which is trained using the $2017$ \emph{Common Object in COntext} (COCO) dataset comprising $118$K image samples with $80$ labels. The input size of model is $416 \times 416$. 
For testing, we randomly select $500$ samples among $5$K validation samples in \cite{COCODataset}. The number of layers in the YOLO model is $106$, divided into FN from layers $1$ to $75$ and IN from layers $76$ to $106$. The number of feature scales is $3$, providing $3$ classes of inference services, namely, a class-$k$ inference using features of scales from $1$ to $k$, where $k\in\{1,2,3\}$.
\subsection{Performance Metrics}
We measure three performance metrics, each of which the definition and evaluation methods are explained below. 

\subsubsection{Communication Load} Based on the concerned Yolo v3 settings, the layer indices corresponding to $3$ scales of features are $\mathbf{f}=[37, 62, 75]$. The resultant computation load of class-$k$ inference is computed by summing up the output data sizes of layers from  $\mathbf{f}(1)$ to $\mathbf{f}(k)$.  

\subsubsection{Computation Load} Denote $\ell_j$ the computation load (in FLOPs) for layer $j$, which can be computed as
\begin{align}
        \ell_j = k_j^2 m_j^2 h_{j-1} h_j, 
\end{align}
where $k_j$, $m_j$, and $h_j$ represents the $j$-th convolution layer's filter size, output size, and channel number, respectively. The resultant computation loads of class-$k$ inference $L(k)$ is 
\begin{align}
    L(k)=\sum_{j=1}^{\mathbf{f}(k)} \ell_j+\sum_{j=\mathbf{i}(k) }^{\mathbf{i}(k)+6}\ell_j,
\end{align}
where the first and second terms represent the computation loads of FN and IN, respectively. Here, each IN comprises $7$ consecutive layers whose starting index is $\mathbf{i}=[76,88,100]$.    
\subsubsection{AI Quality} AI quality can be represented by the precision defined as the probability of detecting objectives correctly. \emph{Mean average precision} (mAP) is the expected precision averaged over objects annotated by different labels. Among several mAP computation methods in the literature, we adopt the technique in \cite{ObjectDetectionPerformanceMetrics}, which is widely used in many object detection applications.
    
\begin{figure*}
\centering
\includegraphics[width=18cm]{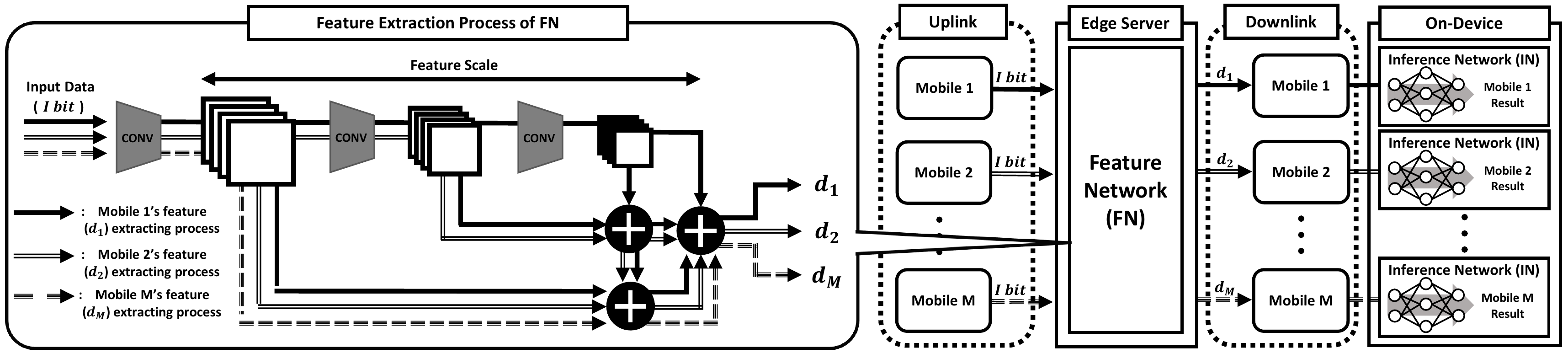}
\caption{Schematic architecture of FHEI, comprising a single edge server and multiple mobiles. The edge server operates FN to extract different scales of features based on the principle of feature hierarchy, and each mobile inputs the extracted feature into the corresponding IN for its corresponding inference~service.}\label{SystemModel}
\end{figure*}
   
\subsection{Observations and Insights}
Fig. \ref{Fig:Service_Quality_Validation} represents the relation among communication load (in MBytes), computation load (in BFLOPs), and AI quality (in mAP) when different classes of inferences are considered. Several interesting observations are made as follows, motivating us to design our system model and formulate the problem introduced in the sequel.

\subsubsection{Effect of a Different Class Inference} A higher class inference results in heavier communication \& computation loads and better AI quality.  

\subsubsection{Linearity w.r.t. Communication Load} Through a linear regression, computation load and AI quality tend to be linearly increasing as a communication load becomes heavier due to a higher class of inference. In other words, a communication load can be interpreted as a controllable variable to adjust both computation load and AI quality.  

\subsubsection{Feature Network's Computation Load Bias}\label{Section:Bias} As shown in Fig. \ref{Exp_Graph 1-1}, FN's computation load is positively biased when the corresponding communication load is the minimum, i.e. $2$ MBytes, which is a baseline computation load to initiate FN. On the other hand, IN's computation load is~unbiased.

\section{Feature Hierarchical Edge Inference:\\ Architecture and Problem Formulation
}\label{System Model}

Prompted by Sec. \ref{Sec: Feasibility}, we propose FHEI to control multiple mobiles' AI qualities. To this end, we firstly introduce the system architecture of FHEI.
Next, several key performance metrics are explained. Last, the optimization problem maximizing the sum of each mobile's AI quality is formulated. 

\subsection{Architecture}
Consider a wireless network comprising $M$ mobiles, denoted by $\mathbb{M}=\{1,\cdots M\}$, and an AP linked to an edge server (see Fig. \ref{SystemModel}). Each mobile attempts to run a DNN-based inference program by the aid of the edge server. To this end, the proposed FHEI splits the computation loads between the edge server and the mobiles, as explained below.

\subsubsection{Edge Inference}
Consider a DNN-based inference program, denoted by $\mathcal{P}_m$, which can be divided into FN $\{\mathcal{F}_m\}$ and IN $\{\mathcal{I}_m\}$. The mobile $m$’s raw data is fixed to certain $I$ Bytes   \footnote{Every mobile's raw data size is assumed to be constant, since it is resized depending on a DNN's predefined input format.}. 
The relation among $\mathcal{P}_m$,  $\mathcal{F}_m$, and $\mathcal{I}_m$ is given as
\begin{align}
\mathcal{P}_m(\boldsymbol{a}_m)
=\mathcal{I}_m(\mathcal{F}_m(\boldsymbol{a}_m))=\mathcal{I}_m(\boldsymbol{b}_m),
\end{align}
where $\boldsymbol{a}_m$ and $\boldsymbol{b}_m$ are vectors representing input and feature for mobile $m$'s program, respectively.

For an effective FN operation, a feature hierarchy-based unified FN can be installed at the edge server, denoted by $\mathcal{F}$, which can cover all mobiles' FNs $\{\mathcal{F}_m\}$, i.e., $\mathcal{F}_m\subset \mathcal{F}$ for all $m\in \mathbb{M}$. Specifically, the unified FN $\mathcal{F}$ follows a feature hierarchy architecture, divided into two parts. The first part includes base layers to initiate the feature extraction. The second one includes hierarchical feature extraction layers such that the scale-$k$ feature with the size of $d^{(k)}$ is extracted from the corresponding layer.
The unified FN $\{\mathcal{F}\}$ can operate as mobile $m$'s FN by extracting the scales of features from $1$ to $k^*(m)$ defined~as
\begin{align}
    k^*(m)=\inf_{k}\left\{d_m\leq \sum_{\ell=1}^{k}d^{(\ell)}\right\},
\end{align}
where $d_m$ represents the size of mobile $m$'s extracted feature.  

Following the observations in Sec. \ref{Section:Bias} and computation model in \cite{Prior1}, the resultant computation load for mobile $m$'s feature extraction (in FLOPs) is given as 
 \begin{align}\label{computation load fn}
    L^{\mathrm{(FN)}}_m= L_0 + c_1 \sum_{\ell=1}^{k^*(m)}d^{(\ell)}\geq L_0 + c_1 d_m,
\end{align}
where $L_0$ (FLOPs) is base layers' computation load and $c_1$ (FLOPs/Byte) is constant depending on the concerned FN. For tractability, we assume that possible scales of feature are well fragmented enough to find $k^*(m)$ satisfying $d_m= \sum_{\ell=1}^{k^*(m)}d^{(\ell)}$ and $L^{\mathrm{(FN)}}_m=L_0 + c_1 d_m$, allowing us to use $d_m$ as a control variable of the optimization introduced in the sequel. 

On the other hand, IN $\{\mathcal{I}_m\}$ remains at mobile $m$'s side to facilitate user-customized services. As observed before, the computation load of mobile $m$'s IN $\{\mathcal{I}_m\}$ is unbiased and linearly increasing of the data size $d_m$, given as 
\begin{align}\label{computation load in}
    {L^{\mathrm{(IN)}}_m} &= c_2 d_m, 
\end{align}
where $c_2$ (FLOPs/Byte) is constant.

\subsubsection{Wireless Communication}
The above computation architecture involves both uplink and downlink transmissions by splitting DNN into FN and IN. To this end, frequency bands for uplink and downlink are exclusively used with the fixed bandwidths of $W_U$ and $W_D$, respectively. 

We consider \emph{time division multiple access} (TDMA) to allows multiple mobiles to access the medium simultaneously. Mobile $m$'s uplink and downlink channel gains are denoted by $g_{m, U}$ and $g_{m, D}$, which are assumed to be stationary within the concerned duration of EI. Following Shannon capacity, the uplink and downlink maximum data rates (in bps) become
    $U_{m}= W_U \log_2\left({1 + \frac{g_{m,U}P_U}{N_0W_U}}\right)$ and
    $D_{m}= W_D \log_2\left(1 + \frac{g_{m,D}P_D}{N_0W_D}\right)$,
where $P_U$ and $P_D$ are transmit power of each mobile and AP, and $N_0$ is a noise spectral density (in Watts/Hz). We denote $\alpha_m$ and $\beta_m$ the time portions assigned for mobile $m$'s uplink and downlink transmissions satisfying $\sum_{m\in \mathbb{M}}\alpha_m \leq 1$ and $\sum_{m\in \mathbb{M}}\beta_m \leq 1$. The resultant achievable rates (in bps) are thus given as 
\begin{align}\label{achievable rate}
    {\Lambda_m}= \alpha_m U_{m},\quad 
    {\Gamma_m} = \beta_m D_{m}.  
\end{align}

\subsection{Key Performance Indicators}

\subsubsection{End-to-End Latency} An E2E latency (in sec), denoted by $T_m$, is defined as the duration required to return a mobile $m$'s inference result, which is expressed as the sum of communication and computation latencies, namely,
\begin{align}\label{total latency}
    T_m = {T^{\mathrm{(comm)}}_m} + {T^{\mathrm{(comp)}}_m}. 
\end{align}
First, communication latency, say $T^{\mathrm{(comm)}}_m$, consists of uplink duration to offload mobile $m$'s raw data to the edge server and downlink duration to download the extracted features from the edge server. Given the data sizes of raw data and extracted features, say $I$ and $d_m$, the communication delay is given as
\begin{align}\label{comm latency}
   {T^{\textrm{(comm)}}_m} = \frac{I}{\Lambda_m} + \frac{d_m}{\Gamma_m}, 
\end{align}
where ${\Lambda_m}$ and ${\Gamma_m}$ are mobile $m$'s uplink and downlink achievable rates specified in (\ref{achievable rate}). Second, computation latency, say ${T^{\textrm{(comp)}}_m}$, consists of an edge server's computation duration for the feature extraction and mobile $m$'s computation duration for the inference of the final result. Assuming that the edge server grants its partial computation resource with the speed of $f_m$ (FLOPs/sec) for mobile $m$, the former is given as $\frac{L^{\textrm{(FN)}}_m}{f_m}$. On the other hand, mobile $m$ computation resource with the speed of $q_m$ (FLOPs/sec) can be entirely used to infer its result. The latter then becomes $\frac{L_m^{\textrm{(IN)}}}{q_m}$. The overall computation latency is 
\begin{align}\label{comp latency}
    {T^{\mathrm{(comp)}}_m} = \frac{L^{\textrm{(FN)}}_m}{f_m} + \frac{L^{\textrm{(IN)}}_m}{q_m}.
\end{align}

By plugging (\ref{comm latency}) and (\ref{comp latency}) into (\ref{total latency}), E2E latency for mobile $m$ is given as
\begin{align}\label{E2D_final_def}
    T_m= \frac{I}{\Lambda_m} + \frac{d_m}{\Gamma_m}+\frac{L^{\textrm{(FN)}}_m}{f_m} + \frac{L^{\textrm{(IN)}}_m}{q_m}.
\end{align}

\subsubsection{Mobile Energy Consumption} Each mobile consumes its energy when communicating with the edge server and computing the final inference result, namely,
\begin{align}\label{total power}
    E_m = {E^{\textrm{(comm)}}_m} + {E^{\textrm{(comp)}}_m}.
\end{align}
First, communication energy consumption, say ${E^{\textrm{(comm)}}_m}$, consists of two parts. The first part represents the energy required to transmit mobile $m$'s raw data, which is the product of the transmit power $P_U$ and offloading duration ${\frac{I}{\Lambda_m}}$. On the other hand, the latter represents the energy required to receive the extracted features from the edge server, which is the product of the receive power $\sigma_m$ and receiving duration ${\frac{d_m}{\Gamma_m}}$. We regard $\sigma_m$ as constant $\sigma$ without loss of generality. The overall communication energy consumption is then given as
\begin{align}\label{comm power}
E^{\mathrm{(comm)}}_m= P_U{\frac{I}{\Lambda_m}}+\sigma{\frac{d_m}{\Gamma_m}}.
\end{align}
Second, following the model in \cite{EnergyConsumption}, computation energy consumption, say ${E^{\textrm{(comp)}}_m}$, is  proportional to the product between the square of computation speed $q_m$ and computation load $L^{\textrm{(IN)}}_m$, namely,
\begin{align}\label{comp power}
    E^{\textrm{(comp)}}_m = {\psi}{q_m}^2{L^{\textrm{(IN)}}_m},
\end{align}
where $\psi$ is the coefficient of computing-energy efficiency. 

By plugging \eqref{comm power} and \eqref{comp power} into \eqref{total power}, the overall energy consumption of mobile $m$ is
\begin{align}\label{Energy_Consumption_final_def}
   E_m= P_U{\frac{I}{\Lambda_m}}+\sigma{\frac{d_m}{\Gamma_m}}+{\psi}\left({q_m}\right)^2{L^{\textrm{(IN)}}_m}. 
\end{align}

\subsection{Problem Formulation}\label{Section:ProblemFormulation}

This subsection formulates the problem of maximizing the sum of AI qualities. First, as discussed in Sec. \ref{Sec: Feasibility}, we define an AI quality as a linear function of a mobile's communication load equivalent to the corresponding feature size $d$, namely, 
 \begin{align}\label{Quality_Function}
        \mathcal{U}(d) = \delta_sd,   
\end{align}
where $\delta_s$ is constant depending on the concerned DNN AI model. Next, we introduce the following optimization problem maximizing the sum of each mobile's AI quality: 
\begin{align}\label{ProblemFormulation}\tag{P1}
    &\max_{\boldsymbol{d}, \boldsymbol{\alpha}, \boldsymbol{\beta}, \boldsymbol{f}, \boldsymbol{q}} \sum_{m \in \mathbb{M}}\mathcal{U}(d_m)\\
    &\begin{aligned}
     \textrm{s.t.} \quad & d_{\textrm{min}} \leq \boldsymbol{d} \leq d_{\textrm{max}}, && {\small \textrm{(Feature Size Constraints)}}\nonumber
     \\
     & \boldsymbol{\alpha}^T\boldsymbol{1} \leq 1, \quad \boldsymbol{\beta}^T\boldsymbol{1} \leq 1, && {\small \textrm{(Radio Resource constraints)}}\nonumber\\
     & \boldsymbol{f}^T\boldsymbol{1} \leq F, \quad  \boldsymbol{q}  \preceq \boldsymbol{\bar{q}}, && {\small \textrm{(Computation resource constraints)}}\nonumber\\
     & \frac{1}{M}\sum_{m \in \mathbb{M}}E_m \leq \bar{E}, && {\small \textrm{(Average E2E Energy constraint)}}\nonumber\\
     & \frac{1}{M}\sum_{m \in \mathbb{M}}T_m \leq \bar{T}, && {\small \textrm{(Average E2E Latency constraint)}}\nonumber\\
    \end{aligned}
\end{align}
where $\boldsymbol{1}$ is a column vector whose all components are $1$. Each constraint in \ref{ProblemFormulation} is explained with relevant definitions. 

\subsubsection{Feature Sizes} Denote $\boldsymbol{d}=[d_1, \cdots, d_M]^T$ a vector whose elements are mobiles' feature sizes, each of which is between $d_{\min}$ and $d_{\max}$ under the constraints below.  

\subsubsection{Radio Resources} Denote $\boldsymbol{\alpha}=[\alpha_1,\cdots,\alpha_M]^T$ and $\boldsymbol{\beta}=[\beta_1,\cdots,\beta_M]^T$ vectors representing mobiles' uplink and downlink time portions, respectively, which are exclusively used for each mobile and limit their sums less than $1$.

\subsubsection{Computation Resources}\label{sec: Computaion Resource Constraint} Denote $\boldsymbol{f}=[f_1,\cdots,f_M]^T$ a vector whose $m$-th component, say $f_m$, represents the edge server's computation resource allocated to mobile $m$ (in FLOPs/sec). The sum of every component cannot exceed the maximum capability defined as $F$. Similarly, $\boldsymbol{q}=[q_1,\cdots,q_M]^T$ is defined as a vector whose $m$-th component, say $q_m$, represents mobile $m$'s computation resource (in FLOPs/sec), less than ${Q}_m$, where $\bar{\boldsymbol{q}}=[Q_1,\cdots,Q_M]^T$.  
\subsubsection{E2E Latency and Energy Consumption Constraints}
The  E2E latency and energy consumption, specified in \eqref{E2D_final_def} and \eqref{Energy_Consumption_final_def} respectively, should be on average less than their requirements of $\bar{T}$ (in sec) and $\bar{E}$ (in Joule). 

\section{AI Quality Optimization of \\ Feature Hierarchical Edge Inference}\label{Section: Proposed Algorithm 1}

\subsection{Problem Reformulation and Overview}\label{Sec: Prob_Reformulation}
This section aim at solving Problem \ref{ProblemFormulation}. The main difficulty lies on the non-convexity of \ref{ProblemFormulation} since the average E2E latency and energy constraints include the terms of multiplying the feature size of $d_m$ with other variables representing radio and computation resources. It can be overcome by decomposing \ref{ProblemFormulation} into the following two sub-problems.  

\begin{figure*}[!t]
\begin{align}\label{Lagrangian_Function}
      \mathcal{L}
    &= \frac{1}{M}\sum_{m \in \mathbb{M}}\left(P_U{\frac{I}{\Lambda_m}}+\sigma{\frac{d_m}{\Gamma_m}}+{\psi}\left({q_m}\right)^2{c_2}{d_m}\right)+ \lambda \left(\frac{1}{M}\sum_{m \in \mathbb{M}}\left(\frac{I}{\Lambda_m} + \frac{d_m}{\Gamma_m} + \frac{L_0 + c_1d_m}{f_m} + \frac{c_2d_m}{q_m}\right) - \bar{T}\right)\nonumber\\
    &+ \mu \left(\sum_{m \in \mathbb{M}}\alpha_m - 1\right) + \gamma \left(\sum_{m \in \mathbb{M}}\beta_m - 1\right)+ \theta \left(\sum_{m \in \mathbb{M}}f_m - F\right) + \sum_{m \in\mathbb{M}}\zeta_m\left(q_m - Q_m\right),
\end{align}
\hrulefill
\end{figure*}

\subsubsection{Joint Radio-and-Computation Optimization} The first sub-problem focuses on optimizing radio and computation resources, say $\boldsymbol{\alpha}$, $\boldsymbol{\beta}$, $\boldsymbol{f}$, and $\boldsymbol{q}$, assuming that the feature size $\boldsymbol{d}$ is given. Specifically,  the objective is to minimize the energy consumption conditioned on $\boldsymbol{d}$ by optimizing the above variables  under the constraints of radio and computation resources and the average E2E latency, as stated below. 
\begin{align}\label{subproblem1}\tag{P2}
    &\min_{\boldsymbol{\alpha}, \boldsymbol{\beta}, \boldsymbol{f}, \boldsymbol{q}}\quad \frac{1}{M}\sum_{m \in \mathbb{M}}E_m\\
    &\begin{aligned}
        \text{s.t. } \quad 
        &\frac{1}{M}\sum_{m \in \mathbb{M}}T_m \leq \bar{T}, \nonumber\\ 
        & \boldsymbol{\alpha}^T\boldsymbol{1} \leq 1, \quad \boldsymbol{\beta}^T\boldsymbol{1} \leq 1, \quad 
        \boldsymbol{f}^T\boldsymbol{1} \leq F, \quad  \boldsymbol{q} \preceq \boldsymbol{\bar{q}}. 
    \end{aligned}
\end{align}

\subsubsection{AI Quality Optimization} Given the variables optimized in \ref{subproblem1}, the second sub-problem is to maximize the sum of AI qualities under the constraint of the average energy consumption as follows.  
 \begin{align}\label{subproblem2}\tag{P3}
    &\max_{\boldsymbol{d}}\quad \sum_{m \in \mathbb{M}}\mathcal{U}(d_m)\\
    &\begin{aligned}
        \text{s.t. } \quad & d_{\textrm{min}} \leq \boldsymbol{d} \leq d_{\textrm{max}}, \quad
        \frac{1}{M}\sum_{m \in \mathbb{M}}E_m \leq \bar{E}.\nonumber
    \end{aligned}
\end{align}

Due to the above decomposition, both \ref{subproblem1} and \ref{subproblem2} are convex problems, enabling us to derive their closed form solutions using the optimization theory, introduced in Sec. \ref{subsection:resource allocation} and Sec. \ref{subsection:quality allocation}, respectively. Next, solving \ref{subproblem1} and \ref{subproblem2} in an iterative manner leads to reach a near-to-optimal solution, whose convergence is explained in Sec. \ref{subsection:convergen proof}. 
 
\subsection{Optimal Radio-and-Computation Resource Allocation}\label{subsection:resource allocation}
This section targets to solve \ref{subproblem1}. Define a Lagrangian function $\mathcal{L}$ as \eqref{Lagrangian_Function} shown in the top of the page, where $\lambda$, $\mu$, $\gamma$, $\theta$ and $\boldsymbol{\zeta}=\{\zeta_1, \cdots ,\zeta_N\}$ denotes Lagrange multipliers associated with the average E2E latency, uplink and downlink bandwidth constraints, edge server and device computation resource constraints, respectively. First, using \emph{Karush Kuhn Tucker} (KKT) conditions, the optimal structures of $\boldsymbol{\alpha}$, $\boldsymbol{\beta}$, and $\boldsymbol{f}$ are derived in terms of $\lambda$, $\mu$, $\gamma$, $\theta$, given as 
\begin{align}
    \alpha_m^* &= \sqrt{\frac{I\left(P_U + \lambda^*\right)}{\mu^* M U_{m}}}, \quad
    \beta_m^* = \sqrt{\frac{d_m\left(\sigma + \lambda^*\right)}{\gamma^* M D_m}}, \nonumber\\
    f_m^* &= \sqrt{\frac{\lambda^*\left(L_{0} + c_1 d_m\right)}{M\theta^*}},
\end{align}
whose derivations are omitted due to the page limit. 
It is observed that $\mu^*$, $\gamma^*$, and $\theta^*$ should be strictly positive for feasible $\alpha_m^*$ $\beta_m^*$, and $f_m^*$, respectively. Due to the slackness condition, the equality conditions of corresponding constraints should be satisfied. Plugging the above optimal structure into the equality condition leads to deriving the closed form solutions of $\alpha_m^*$ $\beta_m^*$, and $f_m^*$, summarized below. 
\begin{proposition}[Optimal Uplink \& Downlink Transmissions and Edge Computing]\label{lemma1} \emph{The optimal solutions of $\alpha_m^*$, $\beta_m^*$, and $f_m^*$ for Problem \ref{subproblem1} are given as
\begin{align*}
\alpha_m^*& = \frac{\frac{1}{\sqrt{U_{m}}}}{\sum_{n \in \mathbb{M}}{\frac{1}{\sqrt{U_{n}}}}},\quad 
\beta_m^* = \frac{\sqrt{\frac{d_m}{D_{m}}}}{\sum_{n \in \mathbb{M}}{\sqrt{\frac{d_n}{D_n}}}},\nonumber\\
f_m^*& = F \frac{\sqrt{L_{0} + c_1 d_m}}{\sum_{n \in \mathbb{M}}\sqrt{L_{0} + c_1 d_n}},\nonumber
\end{align*}
which is independent of mobiles' computation speed $\boldsymbol{q}$.  
}
\end{proposition}
The optimal solution in Proposition \ref{lemma1} gives the minimum time required for uplink \& downlink transmissions and FN execution to assign more time for each mobile's local computing of IN. In other words, Problem \ref{subproblem1} becomes infeasible if the resultant duration of Proposition \ref{lemma1}, which is the minimum duration before local computing, exceeds the E2E threshold $\bar{T}$. The subsequent explanation assumes that the duration is less than $\bar{T}$.

\begin{figure*}[t!]
    \centering
        \subfigure[Allocated AI quality for each mobile]{
        \includegraphics[width=8.6cm, height=3.8cm]{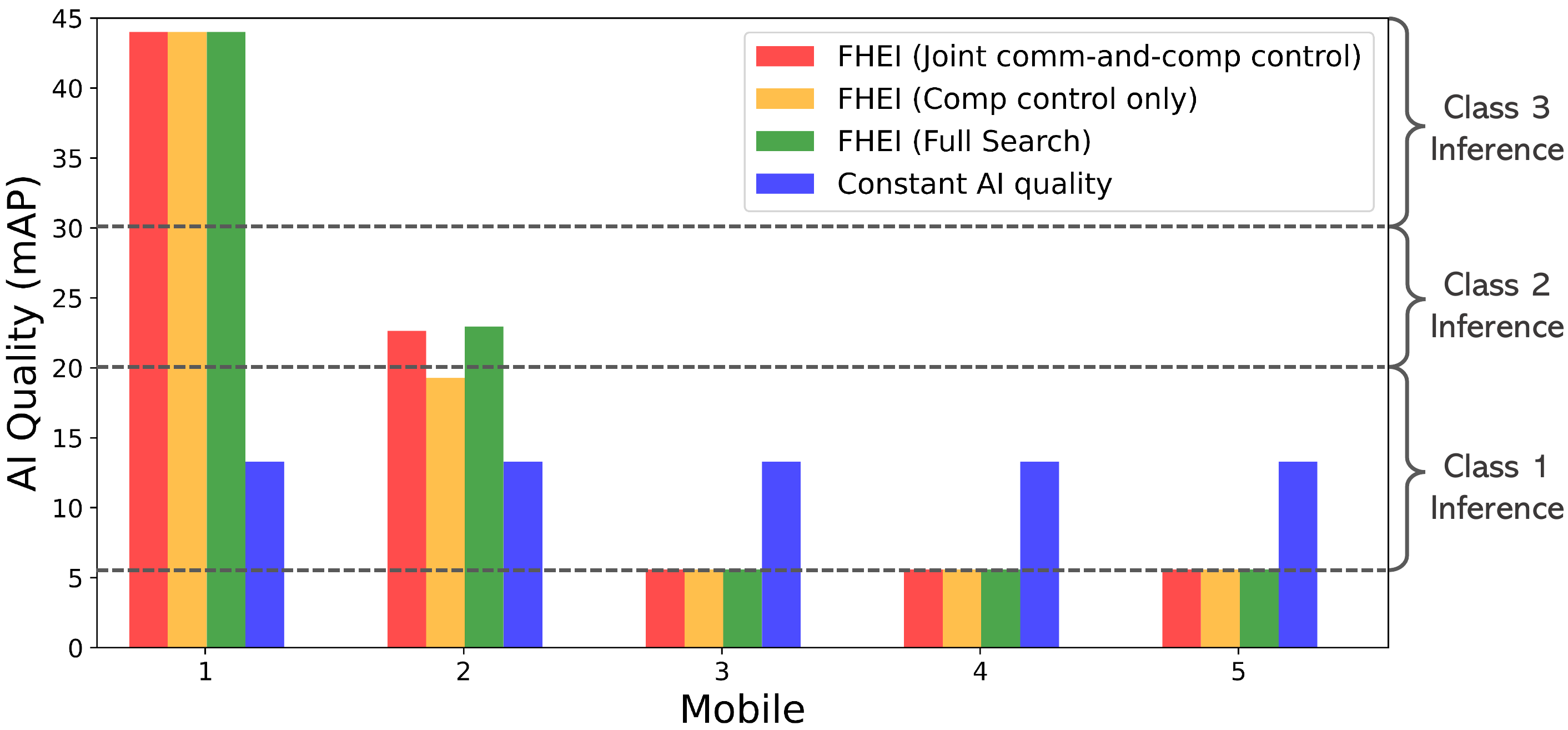}
        \label{Result_Graph 2-1}
        }
        \subfigure[Sum AI quality vs. Number of mobiles]{
        \includegraphics[width=7.8cm,  height=3.8cm]{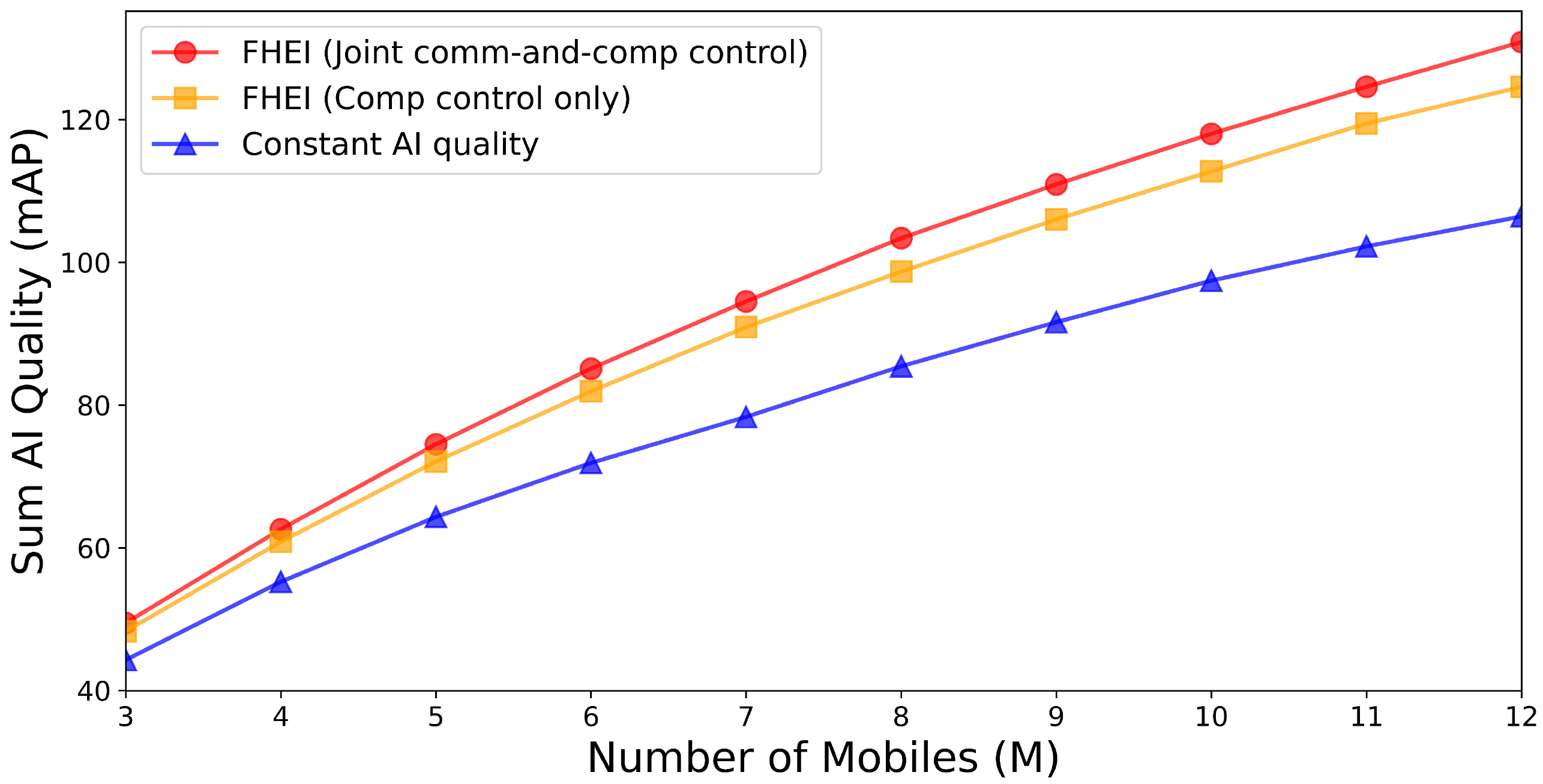}
        \label{Result_Graph 2-2}
        }
     \caption{The performance of FHEI (a) Individual AI quality for each mobiles. The uplink and downlink channel gains of each mobile are given as $[0.05, 1.32, 1.95, 4.63, 3.43]$ and $[5.17, 1.66, 1.51, 0.62, 1.14]$, respectively. (b) Average sum AI quality versus the number of mobiles. The uplink and downlink channel gains follow an independent gamma distribution with unit mean and scale factor of $3$. Other parameters are specified in Table \ref{table_simulation_parameter}.
     }\label{Fig:4 Result Graph}
\end{figure*}

Next, the KKT condition associated with $q_m$, say $\frac{\partial \mathcal{L}}{\partial q_m}=0$, can be manipulated as
\begin{align}
        2 \psi q_m c_2 d_m + \zeta_m - \lambda c_2 \frac{d_m}{q_m^2} = 0,\label{optimal structure for q}
\end{align}
whose closed-form solution is given as
\begin{align}
    q_m^* = \begin{cases} \quad \sqrt[3]{\frac{\lambda^*}{2\psi}} & \textrm{if $\zeta_m = 0$},\\
    \qquad Q_m &  \textrm{if $\zeta_m > 0$},\end{cases}\label{q diviede by zeta}
\end{align}
where mobile $m$'s maximum computation speed $Q_m$ is specified in Sec. \ref{sec: Computaion Resource Constraint}. Noting that the optimal Lagrange multiplier $\lambda^*$ should be strictly positive for non-negative $f_m^*$, the corresponding constraint, say the average E2E delay condition,  satisfies the equality due to the slackness condition, namely, 
\begin{align}
\sum_{m \in \mathbb{M}}\frac{c_2d_m}{q_m^*}=M\bar{T} \!\!-\!\! \sum_{m \in \mathbb{M}}\left(\frac{I}{\Lambda_m^*} \!\!+\!\! \frac{d_m}{\Gamma_m^*} \!\!+\!\! \frac{L_0 + c_1d_m}{f_m^*}\right).  \label{time_const_redefine}
    \end{align}
Plugging \eqref{q diviede by zeta} into \eqref{time_const_redefine} leads to the following proposition. 
\begin{proposition}[Optimal Local Computing]\label{lemma2}\emph{The optimal solution of $q_m^*$ for Problem \ref{subproblem1} is
\begin{align}
    q_m^* = \min\left\{ \sqrt[3]{\frac{\lambda^*}{2\psi}}, Q_m\right\},
\end{align}
where $\lambda^*$ satisfies \eqref{time_const_redefine}.
}
\end{proposition}

\subsection{Optimal AI Quality}\label{subsection:quality allocation}
Noting that Problem \ref{subproblem2} 
is a linear optimization, we can solve it by a well-known greedy algorithm \cite{greedyalgorithm} whose detailed process is omitted due to the page limit.

\subsection{Convergence}\label{subsection:convergen proof}
This section discusses the convergence to a near-to-optimal solution for \ref{ProblemFormulation} by solving Problems \ref{subproblem1} and \ref{subproblem2} attractively. As mentioned before, 
\ref{subproblem1} is optimized based on \ref{subproblem2}'s optimization result, leading to a monotone increasing of the sum of AI qualities until the optimized solution of \ref{subproblem2} is feasible in \ref{subproblem1}. In other words,  the iteration is stopped when \ref{subproblem2}'s solution becomes infeasible in \ref{subproblem1}.

Through extensive numerical studies, one observes that the algorithm is sometimes terminated before reaching an optimal solution especially when the increase of the feature data size $\boldsymbol{d}$ is significant compared with the previous round. It is overcome by adding a constraint limiting the maximum increment per one round in \ref{subproblem2}, given as
\begin{align}\label{NewConstraint}
    \sum_{m \in \mathbb{M}}d_m \leq M \min\{d_{\min} \eta^{k-1}, d_{\max} \}, 
\end{align}
where $\eta>1$ is the increment ratio and $k$ is the round number. Note that \eqref{NewConstraint} is a linear constraint, and \ref{subproblem2} with \eqref{NewConstraint} is solvable using the same method specified in Sec. \ref{subsection:quality allocation}. We set $\eta=1.01$, whose effectiveness is verified in the following~section.

\section{Numerical Results and Concluding Remarks}
This section represents simulation results to verify the effectiveness of FHEI on AI quality control. The concerned parameters are based on the experiment using YOLO v3 in Sec. \ref{Sec: Feasibility},   summarized 
in Table~\ref{table_simulation_parameter}. We consider two benchmarks: constant AI quality and FHEI with computation resource optimization only. For the first benchmark, every mobile's AI quality is fixed but optimized under the same constraints as the proposed FHEI. For the second benchmark, uplink and downlink resources are allocated according to a channel inversion~algorithm, while computation resources are optimized following the method in~Sec. \ref{Section: Proposed Algorithm 1}. 

Fig. \ref{Fig:4 Result Graph} compares FHEI with the above benchmarks. In Fig. \ref{Result_Graph 2-1}, each mobile's optimized AI quality is represented under the given channel conditions specified in the caption. The dotted parallel lines show class inference boundaries when applying the YOLO v3. Several key observations are made. First, the proposed algorithm achieves a near-to-optimal performance. Second, compared with the first benchmark, FHEI can differentiate each user's AI quality depending on a downlink channel state, thereby increasing the entire AI quality. Third, the proposed joint communication-and-computation design outperforms the second benchmark, verifying the validity of the optimization in Sec. \ref{Section: Proposed Algorithm 1}. Last, the gap between the two becomes significant as more mobiles exist (see Fig.~\ref{Result_Graph 2-2}).   

This work focuses on leveraging feature hierarchy to enable AI quality control on EI architecture. On the other hand, several other directions exist for future work, such as applying feature hierarchy to edge learning systems and subsequent resource optimizations.

\begin{table}[]
\centering
\caption{Simulation Settings}
\label{table_simulation_parameter}
\resizebox{\columnwidth}{!}{%
\begin{tabular}{c c c}
\noalign{\smallskip}\noalign{\smallskip}\hline\hline
Notation & Description & Value \\
\hline
$M$                 & \# of mobiles & $5$ \\
$P_U$               & Uplink transmit power                                 & $0.1$W\\
$\sigma$            & Downlink receive power                                & $0.01$W\\
$W_U$               & Uplink  bandwidth                                    & $20$ MHz\\
$W_D$               & Downlink bandwidth                                   & $160$ MHz\\
$F$                 & Edge server computation resource                      & $20$ TFLOPS\\
$Q_m$               & Mobile $m$'s computation resource                     & $1.5 \sim 4.5$ GFLOPS\\
$I$                 & Data size of raw data                                 & $100$ Kbytes\\
$d_{\textrm{min}}$    & Minimum feature size                                  & $2.8$ Mbytes\\
$d_{\textrm{max}}$    & Maximum feature size                                  & $6$ Mbytes\\
$c_1$               & Coefficient of FN                                     & $12.59 \times 10^3$ FLOPs/Byte\\
$c_2$               & Coefficient of IN                                     & $5.664 \times 10^3$ FLOPs/Byte\\
$L_0$               & Constant computation load of FN                       & $14.527$ BFLOPs\\
$\bar{E}$           & Average E2E energy constraint                        & $5$ J\\
$\bar{T}$           & Average E2E latency constraint                       & $15$ sec\\
$\psi$              & Energy efficiency coefficient                         & $10^{-28}$ $\textrm{FLOPs}^3 \times \textrm{J}$/$\textrm{sec}^2$\\
$\delta_s$          & Coefficient of quality function                       & $12$ $\textrm{mAP}/\textrm{Mbyte}$\\
\hline
\hline
\end{tabular}%
}
\end{table}

\bibliographystyle{IEEEtran}
\bibliography{main2}

\begin{thebibliography}{10}
\providecommand{\url}[1]{#1}
\csname url@samestyle\endcsname
\providecommand{\newblock}{\relax}
\providecommand{\bibinfo}[2]{#2}
\providecommand{\BIBentrySTDinterwordspacing}{\spaceskip=0pt\relax}
\providecommand{\BIBentryALTinterwordstretchfactor}{4}
\providecommand{\BIBentryALTinterwordspacing}{\spaceskip=\fontdimen2\font plus
\BIBentryALTinterwordstretchfactor\fontdimen3\font minus
  \fontdimen4\font\relax}
\providecommand{\BIBforeignlanguage}[2]{{%
\expandafter\ifx\csname l@#1\endcsname\relax
\typeout{** WARNING: IEEEtran.bst: No hyphenation pattern has been}%
\typeout{** loaded for the language `#1'. Using the pattern for}%
\typeout{** the default language instead.}%
\else
\language=\csname l@#1\endcsname
\fi
#2}}
\providecommand{\BIBdecl}{\relax}
\BIBdecl

\bibitem{xu2018nature}
J.~Shao and J.~Zhang, ``Communication-computation trade-off in
  resource-constrained edge inference,'' \emph{IEEE Commun. Mag.}, vol.~58,
  no.~12, pp. 20--26, 2020.

\bibitem{Prior1}
Z.~Lin, S.~Bi, and Y.-J.~A. Zhang, ``{Optimizing AI service placement and
  resource allocation in mobile edge intelligence systems},'' \emph{IEEE Trans.
  Wireless Commun}, vol.~20, no.~11, pp. 7257--7271, 2021.

\bibitem{Prior2}
Z.~Liu, Z.~Wu, C.~Gan, L.~Zhu, and S.~Han, ``{Datamix: Efficient
  privacy-preserving edge-cloud inference},'' in \emph{European Conference on
  Computer Vision (ECCV)}.\hskip 1em plus 0.5em minus 0.4em\relax Springer,
  2020, pp. 578--595.

\bibitem{distributed_learning_survey}
M.~Chen, D.~G{\"u}nd{\"u}z, K.~Huang, W.~Saad, M.~Bennis, A.~V. Feljan, and
  H.~V. Poor, ``{Distributed learning in wireless networks: Recent progress and
  future challenges},'' \emph{IEEE J. Sel. Areas Commun.}, 2021.

\bibitem{Task_Orient_EI_MP}
J.~Shao, Y.~Mao, and J.~Zhang, ``{Task-oriented communication for multi-device
  cooperative edge inference},'' \emph{IEEE Trans. Wireless Commun}, 2022.

\bibitem{Branchynet}
S.~Teerapittayanon, B.~McDanel, and H.-T. Kung, ``{Branchynet: Fast inference
  via early exiting from deep neural networks},'' in \emph{Proc. 23rd Int.
  Conf. Pattern Recognit. (ICPR)}.\hskip 1em plus 0.5em minus 0.4em\relax IEEE,
  2016, pp. 2464--2469.

\bibitem{Prior7}
E.~Li, L.~Zeng, Z.~Zhou, and X.~Chen, ``{Edge AI: On-demand accelerating deep
  neural network inference via edge computing},'' \emph{IEEE Trans. Wireless
  Commun}, vol.~19, no.~1, pp. 447--457, 2019.

\bibitem{FeaturePyramid}
T.-Y. Lin, P.~Doll{\'a}r, R.~Girshick, K.~He, B.~Hariharan, and S.~Belongie,
  ``{Feature pyramid networks for object detection},'' in \emph{Proc. IEEE
  Conf. Comput. Vis. Pattern Recognit. (CVPR)}, 2017, pp. 2117--2125.

\bibitem{U-Net}
O.~Ronneberger, P.~Fischer, and T.~Brox, ``U-net: Convolutional networks for
  biomedical image segmentation,'' in \emph{Proc. Int. Conf. Medical Image
  Comput. Comput.-Assisted Intervention}.\hskip 1em plus 0.5em minus
  0.4em\relax Springer, 2015, pp. 234--241.

\bibitem{LaplacianPyramid}
G.~Ghiasi and C.~C. Fowlkes, ``Laplacian pyramid reconstruction and refinement
  for semantic segmentation,'' in \emph{Proc. Eur. Conf. Comput. Vis.}\hskip
  1em plus 0.5em minus 0.4em\relax Springer, 2016, pp. 519--534.

\bibitem{YOLO}
J.~Redmon, S.~Divvala, R.~Girshick, and A.~Farhadi, ``{You only look once:
  Unified, real-time object detection},'' in \emph{Proc. IEEE Conf. Comput.
  Vis. Pattern Recognit. (CVPR)}, Jun. 2016, pp. 779--788.

\bibitem{COCODataset}
T.-Y. Lin, M.~Maire, S.~Belongie, J.~Hays, P.~Perona, D.~Ramanan,
  P.~Doll{\'a}r, and C.~L. Zitnick, ``Microsoft coco: Common objects in
  context,'' in \emph{Proc. 13th Eur. Conf. Comput. Vis}.\hskip 1em plus 0.5em
  minus 0.4em\relax Springer, 2014, pp. 740--755.

\bibitem{ObjectDetectionPerformanceMetrics}
R.~Padilla, S.~L. Netto, and E.~A. Da~Silva, ``A survey on performance metrics
  for object-detection algorithms,'' in \emph{Proc Int. Conf. Syst., Signals
  Image Process.}\hskip 1em plus 0.5em minus 0.4em\relax IEEE, 2020, pp.
  237--242.

\bibitem{EnergyConsumption}
Y.~Wang, M.~Sheng, X.~Wang, L.~Wang, and J.~Li, ``Mobile-edge computing:
  Partial computation offloading using dynamic voltage scaling,'' \emph{IEEE
  Trans. Commun}, vol.~64, no.~10, pp. 4268--4282, 2016.

\bibitem{greedyalgorithm}
J.~Edmonds, ``Matroids and the greedy algorithm,'' \emph{Mathematical
  programming}, vol.~1, no.~1, pp. 127--136, 1971.

\end{thebibliography}

\end{document}